\documentclass[aps,twocolumn,prd,showpacs,showkeys,preprintnumbers,superscriptaddress,nobibnotes,floatfix,longbibliography,nofootinbib]{revtex4-1}

\pdfoutput=1
\usepackage{amsmath}
\usepackage{amsfonts}
\usepackage{amssymb}
\usepackage{mathrsfs}
\usepackage{graphicx}
\usepackage{color}
\usepackage{hyperref}
\usepackage{multirow}
\usepackage{slashed}
\usepackage{epsf}
\usepackage{mathtools}
\usepackage{amsmath,amssymb}
\usepackage{verbatim}
\usepackage{subcaption}

\usepackage{color}
\usepackage{cancel}
\numberwithin{equation}{section}

\usepackage{amsmath, latexsym, amssymb, graphicx, color, multirow, bm, braket, ascmac, bbm}
\usepackage{hyperref}
\usepackage{relsize}
\usepackage{caption}
\usepackage{subcaption}
\usepackage{dsfont}
\usepackage[table]{xcolor}
\usepackage{tabularx}
\usepackage[T1]{fontenc}
\usepackage{mathtools}
\usepackage[capitalize]{cleveref}
\definecolor{nicered}{rgb}{.7,.1,.1}
\definecolor{nicegreen}{rgb}{.1,.5,.1}
\definecolor{darkblue}{rgb}{0,0,.5}
\hypersetup{colorlinks, citecolor=nicegreen,linkcolor=nicered, urlcolor=darkblue}
\usepackage{multirow}
\usepackage{slashed}
\numberwithin{equation}{section}
\usepackage{verbatim}
\usepackage{cancel}

\begin{document}

\title{
Probing Relatively Heavier Right-Handed Selectron at the CEPC, $\rm\bf {FCC_{ee}}$ and ILC 
}

\author{Waqas Ahmed}
\affiliation{School of Mathematics and Physics, Hubei Polytechnic University, Huangshi 435003,China }

\author{Imtiaz Khan}
\affiliation{CAS Key Laboratory of Theoretical Physics, Institute of Theoretical Physics, Chinese Academy of Sciences, Beijing 100190, China}
\affiliation{School of Physical Sciences, University of Chinese Academy of Sciences, No. 19A Yuquan Road, Beijing 100049, China}

\author{Tianjun Li}
\affiliation{CAS Key Laboratory of Theoretical Physics, Institute of Theoretical Physics, Chinese Academy of Sciences, Beijing 100190, China}
\affiliation{School of Physical Sciences, University of Chinese Academy of Sciences, No. 19A Yuquan Road, Beijing 100049, China}

\author{Shabbar Raza}
\affiliation{Department of Physics, Federal Urdu University of Arts, Science and Technology, Karachi 75300, Pakistan}
\author{Wenxing Zhang}
\affiliation{Tsung-Dao Lee Institute and School of Physics and Astronomy, Shanghai Jiao Tong University, 800 Dongchuan Rd., 
Minhang, Shanghai 200240, China}



\begin{abstract}

We employ the low energy Minimal Supersymmetric Standard Model (MSSM) to explore the parameter space 
associated with $Z$-pole and Higgs-pole solutions. Such parameter spaces can not only saturate 
the cold dark matter relic density bound within 5$\sigma$ set by the Planck 2018,
but also satisfy the other standard collider mass bounds and B-physics bounds. 
In particular, we show that the right-handed selectron can be light.
Thus, we propose a search for the relatively heavier right-handed selectron at the future lepton colliders 
with the center-of-mass energy $\sqrt{s}=240$ GeV and integrated luminosity 3000 $\rm{fb^{-1}}$
via mono-photon channel: $e^{+}_{R} e^{-}_{R}\rightarrow {\tilde \chi_{1}^{0}(bino)}+{\tilde \chi_{1}^{0}(bino)}+{\gamma}$. We show that for the Z-pole case the right-handed selectron will be excluded up to 180 GeV and 210 GeV 
respectively at 3$\sigma$ and 2$\sigma$, while the right-handed selectron will be excluded up to 140 GeV 
and 180 GeV respectively at 3$\sigma$ and 2$\sigma$  in case of Higgs-pole.

\end{abstract}
\maketitle

\section{Introduction}
Despite the fact that the Suppersymmetric  Standard Models (SSMs) are the best bet for 
the new physics beyond the standard model (BSM) but no concrete evidence has been found.  
The SSMs predict the unification of the gauge couplings of the hypercharge (or say electromagnetic),  weak and  
strong interactions~\cite{gaugeunification,Georgi:1974sy,Pati:1974yy,Mohapatra:1974hk,Fritzsch:1974nn,Georgi:1974my}, 
provide the natural solution to guage hierarchy problem, and have the lightest supersymmetric particle (LSP) as  
a good cold dark matter candidate~\cite{neutralinodarkmatter,darkmatterreviews}. It is also interesting to note 
that the minimal SSM (MSSM) also predicts the mass range of the Higgs bosom [100,135] GeV~\cite{mhiggs}. 
This is why Supersymmetry (SUSY) has been one of the main targets of the searches being done to look for 
the BSM physics at the Large Hadron Collider (LHC). Despite all the efforts 
at the end of the LHC Run-2 and accumulating data of about 140 $\rm{fb^{-1}}$ at center of mass (CM) energy 
 13 TeV, we have no signs of SUSY. The LHC SUSY searches have put strong constraints on the SSMs. For instance, 
 the masses of the gluino, first-two generation squarks, stop, and sbottom must be larger
than about 2.3~TeV, 1.9~TeV, 1.25~TeV, and 1.5~TeV, 
respectively~\cite{ATLAS-SUSY-Search, Aad:2020sgw, Aad:2019pfy, CMS-SUSY-Search-I, CMS-SUSY-Search-II}.
Thus, at least the colored supersymmetric particles (sparticles) must be heavy around TeV scale.

To escape the LHC SUSY search constraints but remain consistent with various experimental results,
some of us proposed 
the Electroweak Supersymmetry (EWSUSY)~\cite{Cheng:2012np, Cheng:2013hna, Li:2014dna}, 
where the squarks and/or gluinos are
around a few TeV while the sleptons, sneutrinos, Bino and Winos are within about 1 TeV. The
Higgsinos (or say the Higgs bilinear $\mu$ term) can be either heavy or light. Especially, 
the EWSUSY can be realized 
in the Generalized Minimal Supergravity (GmSUGRA)~\cite{Li:2010xr, Balazs:2010ha}.

Apart from high energy collision experiments such as the LHC, the BSM physics is also being probed 
at the low energy. Recently, the Fermi-Lab Collaboration has announced the results for the measurement of
 the anomalous magnetic moment of the muon. 
Combining with the previous results by the Brookhaven National Lab (BNL) experiment, 
we have $4.2 \sigma$ deviation from the SM~\cite{Fermi-Lab} 
\begin{equation}
\label{muon} \Delta a_\mu = a_\mu^{exp}-a_\mu^{th} = (25.1\pm5.9)\times 10^{-10}~.~\,
\end{equation} 
This finding suggests the new physics around 1 TeV and have generated flurry of activity \cite{Crivellin:2021rbq,Endo:2021zal,Iwamoto:2021aaf,Han:2021gfu,Arcadi:2021cwg,Criado:2021qpd,Zhu:2021vlz,Gu:2021mjd,Wang:2021fkn,VanBeekveld:2021tgn,Nomura:2021oeu,Anselmi:2021chp,Yin:2021mls,Wang:2021bcx,Buen-Abad:2021fwq,Das:2021zea,Abdughani:2021pdc,Chen:2021jok,Ge:2021cjz,Cadeddu:2021dqx,Brdar:2021pla,Cao:2021tuh,Chakraborti:2021dli,Ibe:2021cvf,Cox:2021gqq,Babu:2021jnu,Han:2021ify,Heinemeyer:2021zpc,Calibbi:2021qto,Amaral:2021rzw,Bai:2021bau,Baum:2021qzx,Li:2021poy,Zu:2021odn,Keung:2021rps,Ferreira:2021gke,Zhang:2021gun,Ahmed:2021htr}. 

In parallel to LHC and low energy physics experiments for the BSM physics, high energy physics community also have plans to build lepton colliders such as Circular Electro-Positron Collider (CEPC) in China \cite{CEPCStudyGroup:2018rmc,CEPCStudyGroup:2018ghi}, Future Circular Collider ($\rm{FFC_{ee}}$) at CERN \cite{FCC:2018byv,FCC:2018evy},  and International Linear Collider (ILC)~\cite{Behnke:2013xla,Adolphsen:2013kya}. We know that in lepton collider such as $e^{+}e^{-}$, initial states 
are well defined ($E,p$) with known polarization and less background particles are produced after collision . They are ideal machines for high-precision measurements. If we take CEPC as an example, it is a collider with a circumference of 100 km which is designed to operate at center-of-mass energy $\sqrt{s}=$ 240 GeV, 91.2 GeV, and around 160 GeV as Higgs factory, $Z$ factory or $Z$-pole and $W-W$
 threshold scan respectively. It will
produce large samples of Higgs, $W$ and $Z$ bosons to allow precision measurements of
their properties as well as searches for the BSM physics.

The interesting question is whether we can probe the relatively heavier sparticles at the future lepton colliders.
In this article, we study the EWSUSY parameter space via the low scale MSSM boundary conditions. Especially,
 we have the $Z$-pole ($m_{\tilde \chi_{1}^{0}}\approx 1/2 m_{Z}$) and Higgs-pole ($m_{\tilde \chi_{1}^{0}}\approx 1/2 m_{h}$) solutions 
where $m_{\tilde \chi_{1}^{0}}$ is the Bino like LSP neutralino mass. These solutions satisfy the Higgs mass bounds, B-physics bounds, 
and sparticle mass bounds, and have the correct cold dark matter relic density given by Planck 2018. We also present
 a few benchmark points to show the parameter space. In addition to it, making the most of the opportunity of 
the $Z$-pole and Higgs-pole parameter space, for the first time we propose a new search for the relatively heavier 
right-handed selectron 
at future lepton colliders with the center-of-mass energy $\sqrt{s}=240$ GeV and integrated luminosity 3000 $\rm{fb^{-1}}$
via mono-photon channel: $e^{+} e^{-}\rightarrow {\tilde \chi_{1}^{0}(bino)}+{\tilde \chi_{1}^{0}(bino)}+{\gamma}$. 
In this analysis, we consider $e_R^{+}e_R^{-}\rightarrow \nu {\bar {\nu}} \gamma$ as the SM background 
and neglect the events involving $W^{\pm}$ as mediator due to the right-handed selectron search. 
We find that for the $Z$-pole case, the right-handed selectron can be excluded up to 180 GeV and 210 GeV at 3$\sigma$ and 2$\sigma$, 
respectively, while  the right-handed selectron will be excluded up to 140 GeV and 180 GeV at
 3$\sigma$ and 2$\sigma$ in case of Higgs pole, respectively.

The rest of the paper is organized as follows. We describe the input parameter and the ranges for scan in section \ref{sppc},
 display results of scans in section \ref{FS}, discuss our collider study in section \ref{CA}, and conclude  
in section \ref{DC}.



\section{Scanning Procedure and Phenomenological Constraints }
\label{sppc}

\begin{figure*}[ht]
    \centering
        \begin{tabular}{c c}
    \includegraphics[width = 0.5\textwidth]{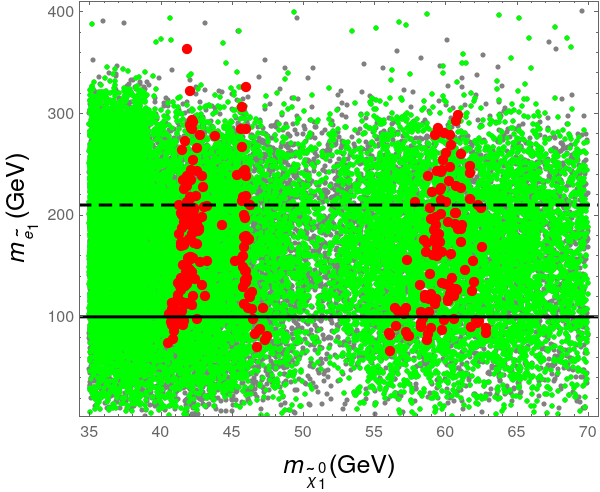} &
        \hspace{-.01cm}\includegraphics[width = 0.5\textwidth]{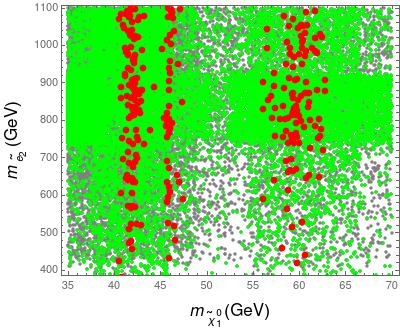}  \\
    \end{tabular}
    \caption{Parameter spaces in the $m_{\tilde{\chi_{1}^{0}}}-m_{\tilde e_{1}}$ and $m_{\tilde{\chi_{1}^{0}}}-m_{\tilde e_{2}}$ planes.
Gray points are consistent with the REWSB and LSP neutralino. Green points satisfy the mass bounds including $m_{h}=125 \pm 3 \,{\rm GeV}$ and the constraints from rare $B$-meson decays. Red points form a subset of green points and satisfy the 5$\sigma$ Planck bounds on dark matter relic density. }
    \label{fig1}
\end{figure*}

\begin{figure*}[ht]
    \centering
        \begin{tabular}{c c}
    \includegraphics[width = 0.49\textwidth]{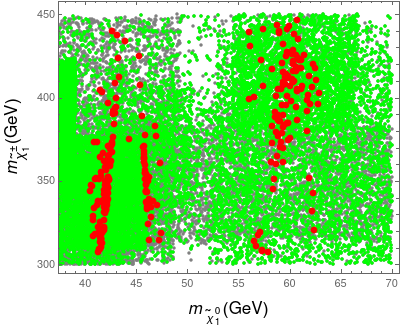} &
        \hspace{-.01cm}\includegraphics[width = 0.52\textwidth]{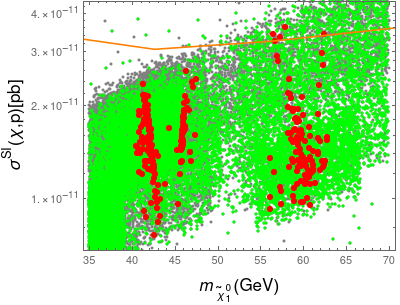}  \\
    \end{tabular}
    \caption{Parameter spaces in the $m_{\tilde{\chi_{1}^{0}}}-m_{\tilde \chi_{1}^{\pm}}$ and $m_{\tilde{\chi_{1}^{0}}}-\sigma^{SI}\left(\chi,p\right)$ planes.
Color coding is same as in Fig.~\ref{fig1}. }
    \label{fig2}
\end{figure*}

We have employed SPheno 4.0.3 package \cite{Porod:2003um, Porod2} {generated} with SARAH 4.13.0 \cite{Staub:2008uz,Staub2} to perform the focused scans to explore the parameter space having Z-resonance and Higgs-resonance solutions. During our focus scan, we therefore adopted the following (universality) conditions to impose on the parameter space of the MSSM at the EW scale:
\begin{gather}
M_{\tilde{f}} \equiv M_{Q_{1,2,3}} = M_{U_{1,2,3}} = M_{D_{1,2,3}}\,, \nonumber \\
T_{\tilde{f}} \equiv T_{\tilde{t}} = T_{\tilde{b}} = T_{\tilde{\tau}}\,, \nonumber
\end{gather}
where $M^2_{Q_{1,2,3}},\,M^2_{U_{1,2,3}}$ and $\,M^2_{D_{1,2,3}}$, are the squared soft masses of the sfermions. The parameter $T_{\tilde{f}}$ corresponds to sfermion trilinear couplings; usually these are taken to be proportional to the Yukawas, such that $T_{\tilde{t}}^{ij} = Y_u^{ij} A_{\tilde{t}},$ where $i,j$ are generation indices. In our numerical code, we fixed all the elements of $T_{\tilde{f}}$ to small values ($1$\,GeV for the diagonal terms and zero otherwise), except for $T_{\tilde{f}}^{(3,3)} = T_{\tilde{t}}^{(3,3)} = T_{\tilde{b}}^{(3,3)} = T_{\tilde{\tau}}^{(3,3)}$, which we left as a free parameter to be scanned over an extended range. We also consider non universal guaginoes and slepton masses.To generate the particle spectrum for a given configuration of the final set of the free parameters,

\begin{center}
	$M_1$\,, $M_2$\,,$M_3$\,, $T_{\tilde{f}}$, $\tan\beta$ \,, $M_{\tilde{f}}$\,, $m_{A}$\,, $\mu$\,,$M^2_{L_{1,2,3}}$\,,$M^2_{E_{1,2,3}}$
\end{center}

In order to calculate $\Omega_{\chi_{1}^{0}}{h^2}$ and other DM observables for each sampled parameter space point, we also produced a {\tt CalcHEP}~\cite{Belyaev:2012qa} model file for the MSSM with {\tt SARAH}, which was then embedded in the public code {\tt MicrOmegas-v5.2.4}~\cite{Belanger:2006is,Belanger:2014vza,Barducci:2016pcb}. 

In scanning the parameter space, we use the SSP \cite{Staub:2011dp}  Mathematica package and link with SPheno and MicrOmegas. The data points collected all satisfy the requirement of REWSB, with the neutralino being the LSP. After collecting the data, we require the following bounds (inspired by the LEP2 experiment) on sparticle masses.\\
\textbf{\underline{LEP constraints:}}
We impose the bounds that the LEP2 experiments set on charged sparticle masses ($\gtrsim 100$ GeV) \cite{Patrignani:2016xqp}.

\textbf{\underline{Higgs Boson mass:}}
The experimental combination for the Higgs mass reported by 
the ATLAS and CMS Collaborations is \cite{Khachatryan:2016vau}
\begin{align}\label{eqn:mh}
m_{h} = 125.09 \pm 0.21(\rm stat.) \pm 0.11(\rm syst.)~GeV .
\end{align}  
Due to the theoretical uncertainty in the Higgs mass calculations in the MSSM -- see {\it e.g.}~\cite{Slavich:2020zjv,Allanach:2004rh} -- we apply the constraint from the Higgs boson mass to our results as: 
\begin{align}\label{eqn:higgsMassLHC}
122~ {\rm GeV} \leq m_h \leq 128~ {\rm GeV}. 
\end{align}

\textbf{\underline{Rare B-meson decays:}} Since the SM predictions are in a good agreement with the experimental results for the rare decays of $B-$meson such as the $B_{s}\rightarrow \mu^{+}\mu^{-}$, $B_{s}\rightarrow X_{s}\gamma$, where $X_{s}$ is an appropriate state including a strange quark, the results of our analyses are required to be consistent with the measurements for such processes. Thus we employ the following constraints from B-physics \cite{CMS:2014xfa,Amhis:2014hma}:

\begin{align}\label{eqn:Bphysics}
1.6\times 10^{-9} \leq ~ {\rm BR}(B_s \rightarrow \mu^+ \mu^-) ~
\leq 4.2 \times10^{-9} ,\\ 
2.99 \times 10^{-4} \leq  ~ {\rm BR}(b \rightarrow s \gamma) ~
\leq 3.87 \times 10^{-4}, \\
0.70\times 10^{-4} \leq ~ {\rm BR}(B_u\rightarrow\tau \nu_{\tau})~
\leq 1.5 \times 10^{-4}.
\end{align}

\textbf{\underline{Current LHC searches:}}
Based on \cite{Aaboud:2017vwy,Vami:2019slp,Sirunyan:2017kqq}, we consider the following constraints on gluino and first/second generation squark masses
\begin{align}
(a) \quad \quad m_{\widetilde g} \gtrsim ~ 2.2 ~ {\rm TeV},\quad \quad m_{\widetilde q} \gtrsim ~ 2 ~ {\rm TeV},
\label{lhc-a}
\end{align}

\textbf{\underline{DM searches and relic density:}} For the discussion on the phenomenology of neutralino DM in our scenario, we impose the following constraint for the LSP relic density, based on {the current measurements of the Planck satellite} \cite{Akrami:2018vks}:
\begin{align}\label{eq:omega}
0.114 \leq \Omega_{\rm CDM}h^2 (\rm Planck2018) \leq 0.126   \; (5\sigma).
\end{align}

We use the current  XENON1T with 2 $t\cdot y$ spin-independent (SI) DM cross section with bounds \cite{Aprile:2015uzo}.  All points lying above these upper bounds have been excluded from the plots .
\section{Results of focused scans}{\label{FS}}

In this section we are focusing on the sparticle spectrum consistent with mass bounds and the constraints discussed above.

In Figure~\ref{fig1}, we display plots in $m_{\tilde{\chi_{1}^{0}}}-m_{\tilde e_{1}}$ and $m_{\tilde{\chi_{1}^{0}}}-m_{\tilde e_{2}}$ planes. Note that here $\tilde e_{1}$ is the right-handed selectron while $\tilde e_{2}$ is left handed selectron. Gray points satisfy the REWSB and the LSP neutralino conditions. Green points satisfy the mass bounds and B-physics constraints. Red points form a subset of green points and satisfy the Planck 2018 bounds (mentioned above) on the relic abundance of the LSP neutralino within $5\sigma$ uncertainty. We see that from left panel the green points are almost everywhere. But when we demand the relic density bound, red solutions appear to be around $m_{\tilde \chi_{1}^{0}}\sim$ 45 GeV and 62 GeV. These solutions represent well known the Z-pole ($m_{\tilde \chi_{1}^{0}}\approx 1/2 m_{Z}$) and the Higgs-pole ($m_{\tilde \chi_{1}^{0}}\approx 1/2 m_{h}$) solutions where two LSP neutralinos annihilate via s-channel exchange of a virtual particle, for example, a Z or Higgs boson and the mass of the exchanged particle closely matches twice the LSP neutralino mass. The solid black horizontal line indicates the LEP bounds on sleptons \cite{LEP:Working,ALEPH:2003acj,DELPHI:2003uqw,L3:1999onh,OPAL:2003wxm} while the dashed black line represents the estimated exclusion limit for right handed selectron(we will discuss it in section~\ref{CA}). Note that though red points appear to be between 80 GeV to 360 GeV in our present scans but we can increase the mass range by keep doing focus scans.  In the left panel we see the similar situation for neutralino mass and we note that $\tilde{e_{2}}$ or the left handed slepton mass can be as heavy as 1100 GeV.

In Figure~\ref{fig2} left panel, we show plots for  $m_{\tilde{\chi_{1}^{0}}}-m_{\tilde \chi_{1}^{\pm}}$ and $m_{\tilde{\chi_{1}^{0}}}-\sigma^{SI}\left(\chi,p\right)$. Color coding is same as in Fig.~\ref{fig1}. In the left panel we see that $m_{\tilde \chi_{1}^{\pm}}$ can be as heavy as 450 GeV for both Z-pole and Higgs-pole scenario but can be made heavy by focused scans. In the right panel display $m_{\tilde{\chi_{1}^{0}}}-\sigma^{SI}\left(\chi,p\right)$
plot. Her orange line represents the current XENON1T with 2 $t\cdot y$ \cite{Aprile:2015uzo} bounds. In this plot, the two dips around 45 GeV and 62 GeV indicate the Z-pole and Higgs-pole solutions. This plot clearly show that almost all of the red points are consistent with the current bounds set by XENON1T with 2 $t\cdot y$ \cite{Aprile:2015uzo}.

\begin{table}[h!]
	\centering
	\scalebox{0.8}{
		\begin{tabular}{lccccc}
			\hline
			\hline

                 & Point 1 1 & Point 2 & Point 3 & Point 4 \\

\hline
$M_{1} $            & 59.67    & 48.3       & 64.38    & 63.99  \\
$M_{2}$             & 650.33    & 641.2      &  486.97 &  494.92 \\
$M_{3}$             & -3720.77  & -2579.8   & -3030  & -4024\\
$M_{\tilde f}$      & 3299.84   & 3456.79   &  3083.03   & 3146.76 \\
$m_{\tilde L}$      & 848.39    & 824.63    &  912.81     &  1005.25\\
$m_{\tilde E^c}$    & 371.121    & 340.65       &  323.66     &  379.464\\
$T_f$           & -3793.68    & -2597.75       &  -2952.52  &  -4429.96\\
$\tan\beta$         & 57.2    & 56.9      & 59.96     &  47.69\\
$\mu$           & 318.24   & 339      & 361.1   & 440.6\\
\hline
$m_h$            & 125.5  & 123     & 124  &  127\\
$m_H$             &4095.9   & 4000    & 3842  & 4093 \\
$m_A$            & 4096   & 4000     & 3842 &  4093\\
$m_{H^{\pm}}$    & 4104.39   & 4041     & 3858  & 4102 \\
\hline
$m_{\tilde{\chi}^0_{1,2}}$
                 & 57, 319 & 46, 339     & 62, 351  & 62, 391\\

$m_{\tilde{\chi}^0_{3,4}}$
                 & 330,693  & 351, 684    & 373, 535   & 417, 548\\

$m_{\tilde{\chi}^{\pm}_{1,2}}$
                 & 319, 692  & 339, 684   & 352, 536  &  392, 549\\
\hline
$m_{\tilde{g}}$  & 3864 & 2872  &  3234 &  4087\\
\hline
$m_{ \tilde{u}_{1,2}}$
                 & 3252, 3434  & 3422, 3542   & 3051, 3195   & 3059, 3283\\
$m_{\tilde{t}_{1,2}}$
                 & 3430, 3440 & 3542, 3552   & 3192, 3202   & 3274, 3288\\
\hline
$m_{ \tilde{d}_{1,2}}$
                 & 3354, 3430 & 3475, 3543    & 3122, 3193  & 3197, 3275 \\
$m_{\tilde{b}_{1,2}}$
                 & 3430, 3441   & 3543, 3553   &  3193, 3203   & 3275, 3285 \\
\hline
$m_{\tilde{\nu}_{1}}$
                 &799 & 785   & 875  & 1014 \\
$m_{\tilde{\nu}_{3}}$
                 &851 & 826    & 911   & 1054 \\
\hline
$m_{ \tilde{e}_{1,2}}$
               & 102, 805  & 120, 791    & 125,881  & 215, 1018 \\
$m_{\tilde{\tau}_{1,2}}$
                & 391, 855   & 364, 831   & 346, 915   &  400, 1057\\
\hline

$\sigma_{SI}({\rm pb})$
                & $2.79\times 10^{-11}$ & $ 1.72\times 10^{-11} $  & $2.33\times 10^{-11}$  &  $1.38 \times 10^{-11}$\\

$\sigma_{SD}({\rm pb})$
                & $ 1.33\times 10^{-5}$ &$ 1 \times 10^{-5} $  & $8.18\times 10^{-6}$ &  $5.03\times 10^{-6}$\\

$\Omega_{CDM}h^{2}$ &0.116  & 0.118  & 0.1257  &  0.1259 \\
\hline
\hline
\end{tabular}
}
\caption{\small{The sparticle and Higgs masses (in GeV units) for the benchmark points.}}
\label{table1}
\end{table}
In order to show the glimpse of the parameter space we display four benchmark points. Point 1 and Point 2 are examples of Z-pole solutions and Point 3 and Point 4 display examples of Higgs-pole solutions.
\section{Collider Analysis}{\label{CA}}

\begin{figure}[ht]
	\begin{subfigure}{.15\textwidth}
		\includegraphics[width=1.0\linewidth]{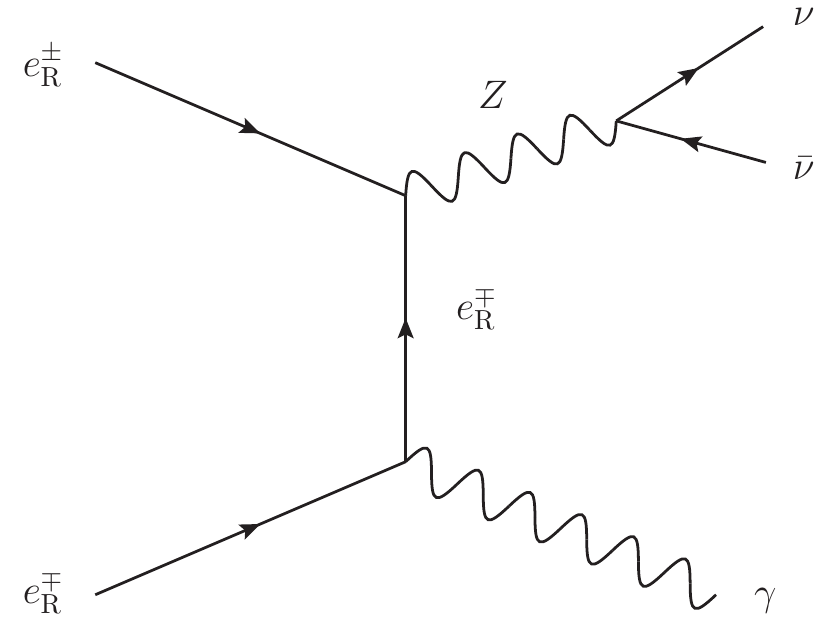}  
		\caption{\label{fig:fd-a}}
	\end{subfigure}
	\begin{subfigure}{.15\textwidth}
		\includegraphics[width=1.0\linewidth]{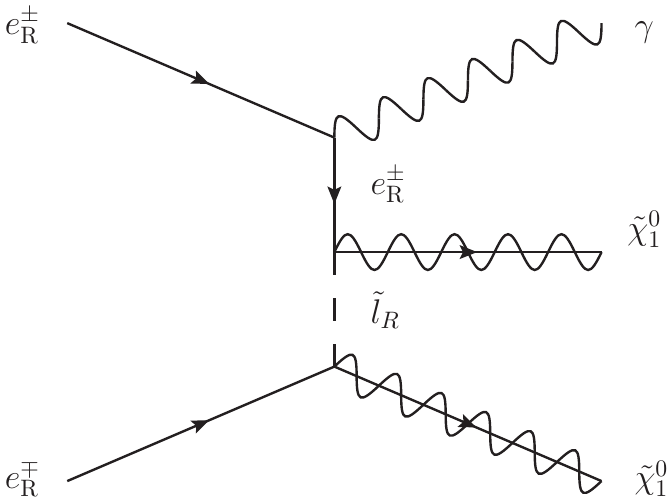}  
		\caption{\label{fig:fd-b}}
	\end{subfigure}
	\begin{subfigure}{.15\textwidth}
		\includegraphics[width=1.0\linewidth]{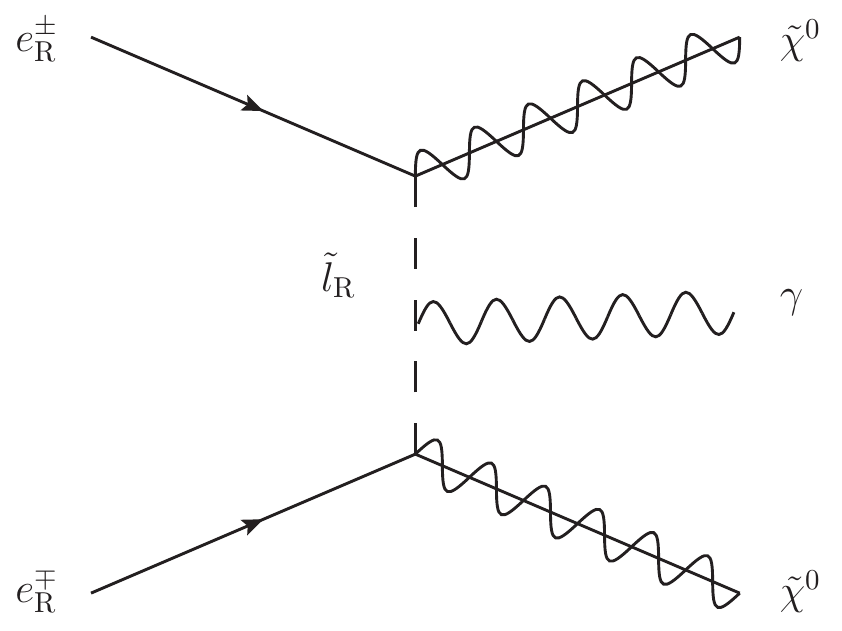}  
		\caption{\label{fig:fd-c}}
	\end{subfigure}
	\caption{The Feynman diagrams for the background and signal events, respectively.}
	\label{fig:feynman}
\end{figure}
In this section, we investigate the lepton collider approach to bino-like LSP via a light right-handed selectron.
We study the final states with missing transverse energy induced by two bino-like LSP plus an emissive mono-photon and focus on the t-channel process mediated by an off-shell right-handed slepton as shown in Fig~\ref{fig:feynman}. 
In principle, such processes can probe slepton mass heavier than the collision energy.
In order to further suppress the background events associated with W-boson process, we consider the polarized electron-positron collision with missing transverse energy induced by two neutralinos and an emissive mono-photon. 
The corresponding background events containing missing energy and mono-photon comes from Drell-Yan process with Z boson decay to two neutrinos. 

In our numerical simulation, Monte Carlo samples of signal and background events are generated by using MadGraph5~\cite{Alwall:2014hca, Frederix:2018nkq} for hard scattering processes,
PYTHIA8~\cite{Cacciari:2011ma} for parton showering and hadronization
and DELPHES 3~\cite{deFavereau:2013fsa} for jet clustering and detector simulation.
We generate the signal events with the collision energy equal to 240 GeV and use the $anti-k_T$ algorithm to do the jet reconstruction with the radius $R=0.4$. The beam polarization is set to be fully right-handed. 
All jets and particles within $|\eta| < 3.0$ will be recorded by simulation, otherwise it will be missed.
In addition, we apply photon isolation techniques, which have been developed to filter out indirect photons that are produced from the fragmentation of quark and gluon partons.
Photon isolation viable is defined as
\begin{equation}
	I(P)=\frac{\mathlarger {\mathlarger{\sum}}_{i\neq P}^{\Delta R < R_0 , p_T^i > p_T^{min}} p_T(i) }{p_T(P)},
\end{equation}
where the denominator stands for the transverse momentum of photon, and the numerator is the sum of transverse momenta above $p_T^{min}$ of all particles that lie within a cone of radius $R_0$ around and except the photon.
In our simulation, we require the isolation viable to be $I(P) < 0.12$ and $R_0 = 0.5$, $p_T^{min}=0.5$ GeV.

The final states is quite simple with only MET and mono-photon to be detected.
Therefore, in order to analyze the signal events, we make use of the following quantities and the threshold is chosen to maximize the Higgs signal significance of the CEPC:

\begin{figure*}[ht]
    \centering
        \begin{tabular}{c}
    \includegraphics[width = 0.5\textwidth]{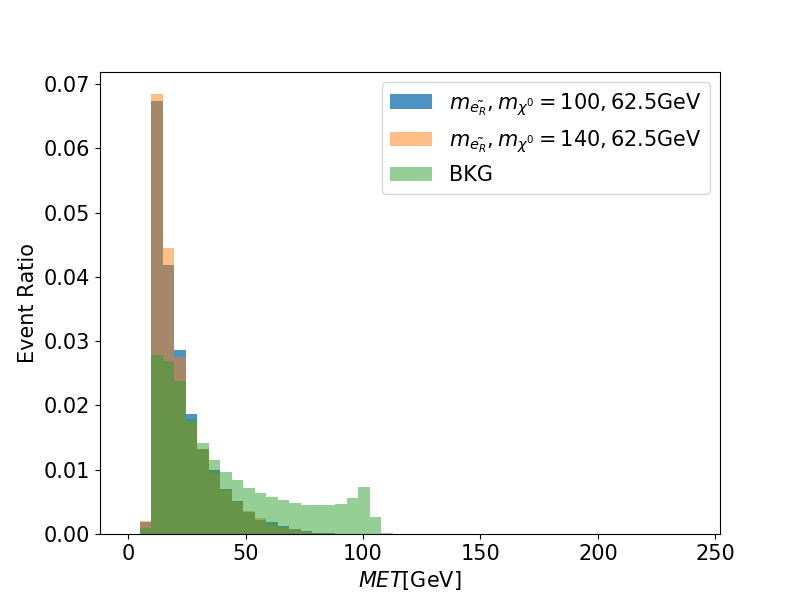} 
       \hspace{-.01cm}\includegraphics[width = 0.5\textwidth]{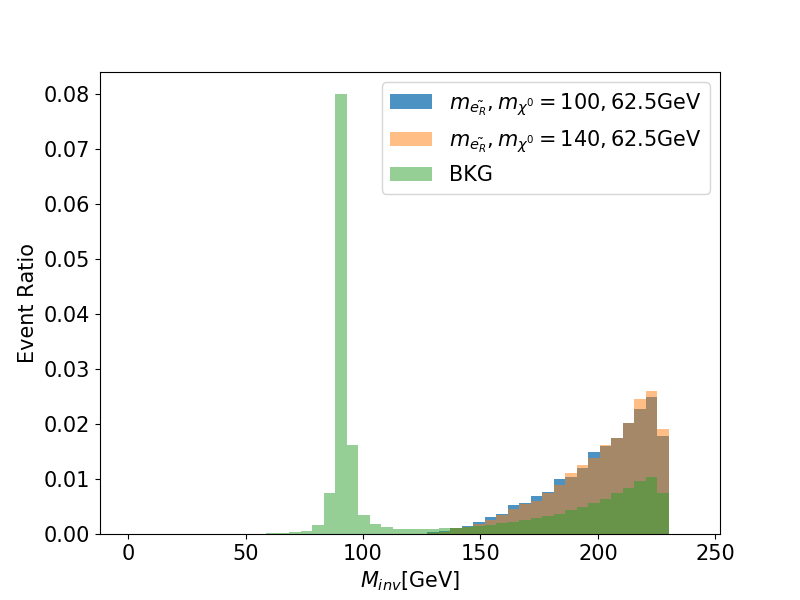}  \\
    \end{tabular}
    \caption{The distribution of missing transverse momentum (left) and the invariant mass of the missing momentum (right) for Higgs-pole case with $m_{\tilde{e}_R}=100$ GeV and $m_{\tilde{e}_R}=140$ GeV.}
    \label{fig4}
\end{figure*}

\begin{itemize}
	\item Missing transverse energy (MET) of all invisible particles: We require $\cancel{E}_{inv} < 80.0$. The distribution of MET is shown in the left panel of Fig.~\ref{fig4}, where the green region stands for the background distribution and the blue/orange one stands for the Higgs-pole case with slepton mass equal to 100 and 140 GeV respectively.
	\item Invariant mass of invisible particles: $m_{inv}=\sqrt{(E_{total}-E_{vis})^2 - p^2_{vis}  }$, where $E_{total}=240$ GeV, $E_{vis}$ the total energy of all visible particles and $p_{vis}$ the visible particles. It should satisfy: $m_{inv} >130.0$ GeV.
	Since for background events, two neutrinos come from an on-shell Z boson, so the invariant mass of two neutrinos is around Z boson mass as shown in the right panel in Fig.~\ref{fig4}. In this panel, the color meaning is same as before, and as we can see, the distribution of Z-pole mass set an effective cut-off between signal events and background events.
\end{itemize}
Note that here we show our study for Higgs-pole only but more or less similar results true for Z-pole solutions.

\begin{figure}[t!]
\includegraphics[width=1.0\columnwidth]{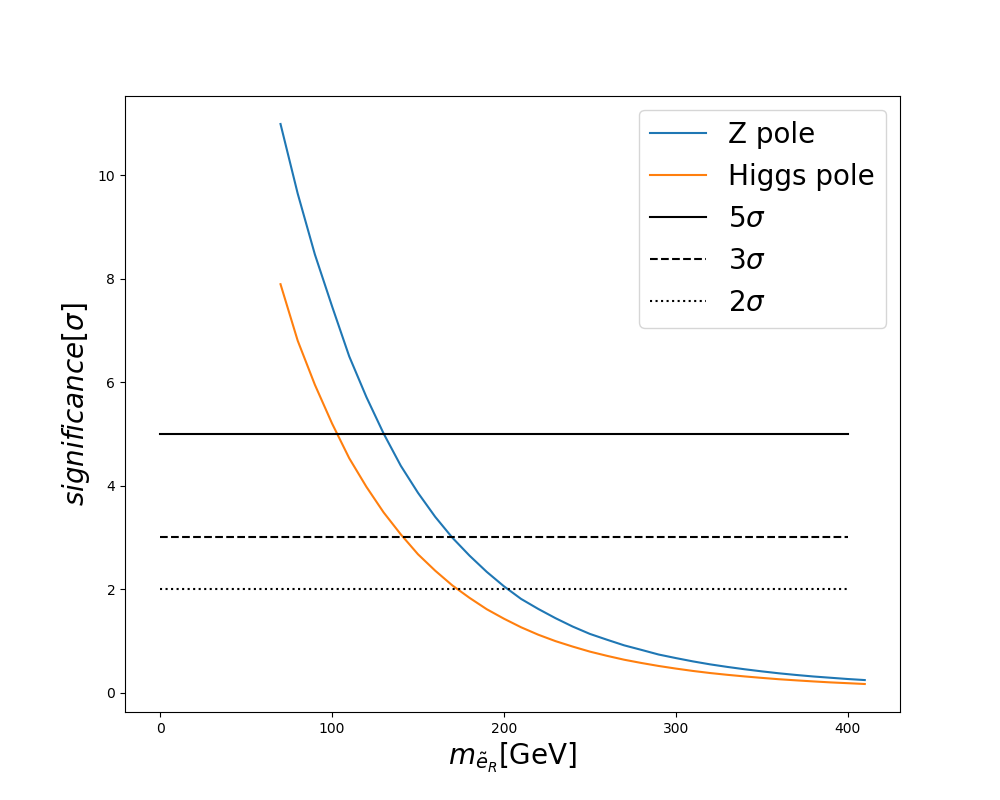}
\caption{
Significance of exclusion ability.}
\label{fig5}
\end{figure}

Significance for signal events is calculated by $\sigma = S/\sqrt{B}$. In Fig.~\ref{fig5}, we show the significance of signal events with the selectron mass ranging from 50 GeV to 400 GeV. 
For the Z-pole case, the right handed selectron will be excluded to 180 GeV and 210 GeV by 3$\sigma$ and 2$\sigma$ respectively.
In the Higgs-pole scenario, the right handed selectron will be excluded to 140 GeV and 180 GeV by 3$\sigma$ and 2$\sigma$ respectively.

\section{Discussion and Conclusion}{\label{DC}}

In the low energy MSSM, we studied the EWSUSY parameter space associated with $Z$-pole and Higgs-pole solutions. 
Such parameter spaces can not only have the correct dark matter relic density within 5 $\sigma$ from the Planck 2018,
but also escape the other standard collider mass bounds and B-physics bounds. 
Especially, the right-handed selectron can be light.
Therefore, we proposed a search for the relatively heavier right-handed selectron at the future lepton colliders 
with the center-of-mass energy $\sqrt{s}=240$ GeV and integrated luminosity 3000 $\rm{fb^{-1}}$
via mono-photon channel: $e^{+}_{R} e^{-}_{R}\rightarrow {\tilde \chi_{1}^{0}(bino)}+{\tilde \chi_{1}^{0}(bino)}+{\gamma}$. 
We showed that for the $Z$-pole case the right-handed selectron can be excluded up to 180 GeV and 210 GeV 
respectively at 3$\sigma$ and 2$\sigma$, while the right-handed selectron can be excluded up to 140 GeV 
and 180 GeV respectively at 3$\sigma$ and 2$\sigma$  in case of Higgs-pole.

\section*{Acknowledgements}

TL is supported by the National Key Research and Development Program of China Grant No. 2020YFC2201504, 
by the Projects No. 11875062, No. 11947302, and No. 12047503 supported 
by the National Natural Science Foundation of China, as well as by the Key Research Program 
of the Chinese Academy of Sciences, Grant NO. XDPB15. 


\end{document}